# A Data Transmission Method Based on Ethernet Physical Layer for Particle Physics Experiment*

HUANG Xi-Ru(黄锡汝)[1,2; 1)] CAO Ping(曹平)[1,2; 2)] ZHENG Jia-Jun(郑佳俊)[1,2]

[1] State Key Laboratory of Particle Detection and Electronics, University of Science and Technology of China, Hefei 230026, China
[2] Anhui Key Laboratory of Physical Electronics, Department of Modern Physics, University of Science and Technology of China, Hefei 230026, China

**Abstract:** Due to the advantages of universality, flexibility and high performance, fast Ethernet is widely used in readout system design of modern particle physics experiments. However, Ethernet is usually used together with TCP/IP protocol stack, which makes it difficult to be implemented because designers have to use operating system to process this protocol. Furthermore, TCP/IP protocol degrades the transmission efficiency and real-time performance. To maximize the performance of Ethernet in physics experiment applications, a data readout method based on physical layer (PHY) is proposed in this paper. In this method, TCP/IP protocol is forsaken and replaced with a customized and simple protocol, which make it easier to be implemented. On each readout module, data from front-end electronics is first fed into an FPGA for protocol processing and then sent out to a PHY chip controlled by this FPGA for transmission. This kind of data path is fully implemented by hardware. While from the side of data acquisition system (DAQ), absence of standard protocol makes the network related applications panic. To solve this problem, in the operating system kernel space, data received by network interface card is drawn away from the traditional flow and redirected to a specified memory space by a customized program. This memory space can be easily accessed by applications in user space. For the purpose of verification, a prototype system is designed and implemented. Preliminary test result shows that this method can meet the requirement of data transmission from readout module to DAQ with good efficiency and simplicity.

## 1 Introduction[1]

In modern particle physics experiment, with the increasing of demands of large amount of electronic channels, massive data and short readout period, there's great challenge for readout system design [1-2]. On the other hand, the newly developed and gradually used waveform digitizing technique also brings enormous pressure for readout system design [3-5]. Readout system needs to concentrate massive data from front-end electronics (FEE) and send to data acquisition system (DAQ) in real-time. Generally, readout system is comprised of some standard crates (e.g. VME) inside which some readout modules as shown in Fig. 1 are settled. Each readout module receives data from FEE and then sends data to the crate controller one by one through the crate backplane bus [2, 6]. Finally, controller module sends the concentrated data to DAQ through Ethernet channel on it.

This readout scheme is a distributed architecture that each crate controller acts as a network node and the DAQ acts as data receiving center. All nodes are connected together by Ethernet. To improve the data transmission performance, Gigabit Ethernet is usually used in this scheme. For the case of experiments with large amount of electronic channels, to balance data transmission performance, designers can adjust the number of readout crates and electronic modules in them.

Unfortunately, there are some weak points in this scheme. Firstly, transmission load balancing gives pressure for readout crates. Readout system designers have to increase either the number of readout creates or the performance of backplane bus and controller. Secondly, combined with TCP/IP protocol stack, the efficiency of Ethernet is degraded seriously. To guarantee the reliability, in TCP/IP protocol, redundant information is encapsulated into each data packet. Obviously, redundant information benefits for transmission reliability but not for efficiency. Actually, in physics experiment applications, because of the simple network connectivity, transmission reliability can be easily guaranteed, while on the contrary, the transmission efficiency needs being paid more attention.

Fig. 2 shows another different readout scheme. In this scheme, each readout module can send data to DAQ separately. There are many advantages compared with the traditional scheme. Firstly, there's no any direct data transmission path between readout module and crate controller. The uploaded data from FEE is transmitted by readout module in parallel. Crates with lower performance backplane bus can be used for various experiments with large or small scale. So this scheme has advantage for system expansibility. Secondly, controller and backplane bus are free from the readout data transmitting flow and switched to the purpose of system control or configuration. In fact, in physic experiment applications, the path of data (received from FEE) uploading differs significant from that of data (received from DAQ) downloading. The data throughput or real-time performance demand of the former is stricter than the latter. In this new scheme, these two kinds of data path are separated and implemented as full duplex formation. Shortly, this new scheme raises no pressure for backplane bus and controller.

---
[1] * Supported by National Natural Science Foundation of China (11005107) and independent project of State Key Laboratory of Particle Detection and Electronics (201301)
1) E-mail: xiru@mail.ustc.edu.cn
2) E-mail: cping@ustc.edu.cn

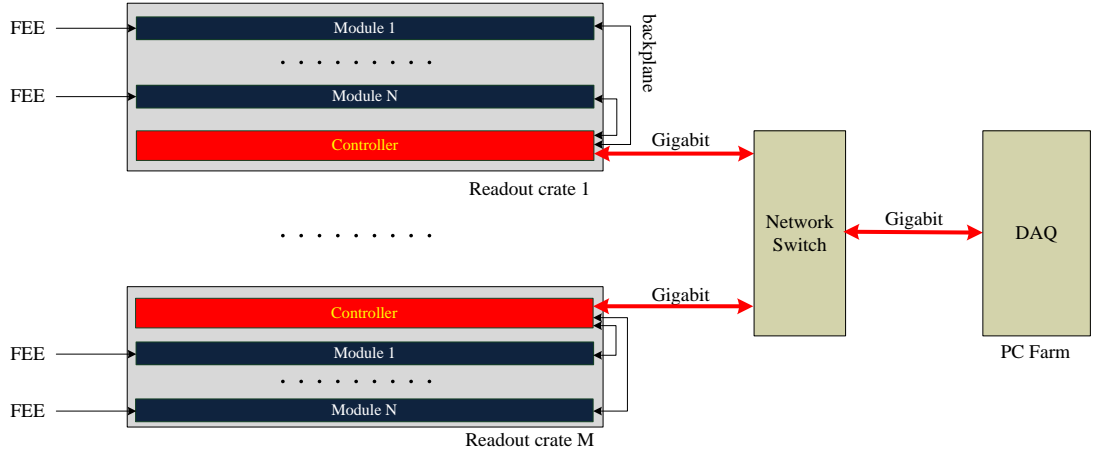

Fig. 1.  (color online) Readout scheme based on crates.

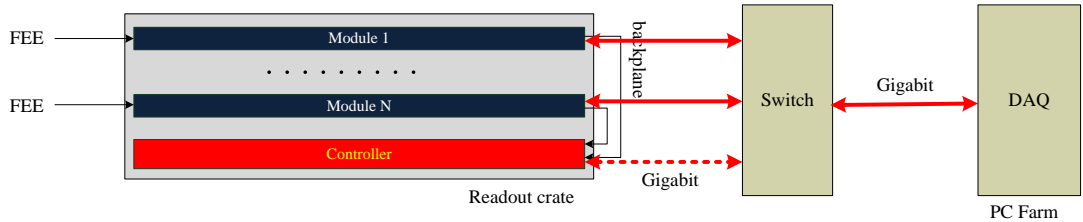

Fig. 2.  (color online) Readout scheme based on individual modules.

## 2 Readout method based on Ethernet PHY

The Ethernet technique plays an important role for realizing this advance readout scheme. As all known, Ethernet is always used combined with TCP/IP protocol. To support TCP/IP protocol while implementing an electronic readout module, usually, a CPU and embedded operating system are utilized [7-8]. Embedded system has disadvantages about cost, complexity and power consumption etc. Some experiments use FPGA logic core to implement this embedded system. But it is hard to achieve good stability and good performance [9].

Fig. 3 shows a new scheme of Ethernet implementation.

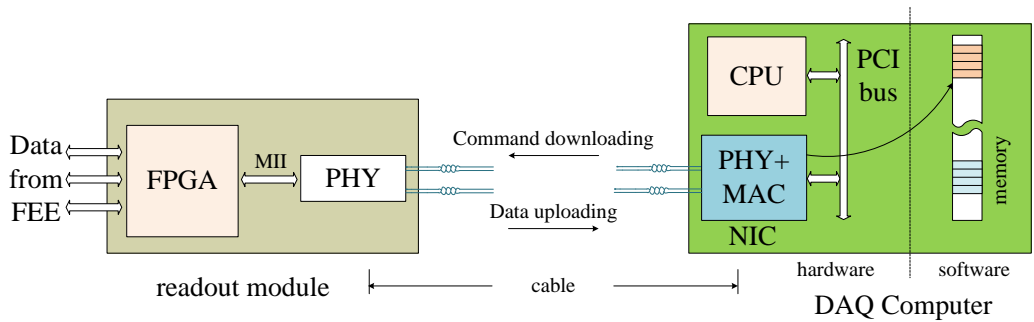

Fig. 3.  (color online) Readout scheme based on Ethernet PHY.

In this scheme, data from FEE is fed into the FPGA on each readout module. According to the processing performance, one readout module can receive data from FEE channels. The FPGA's main task is protocol resolving and data packaging. The re-packaged data is then sent to a standalone PHY chip for transmitting over long cable connected to DAQ computer. This PHY chip is controlled by the FPGA through media independent interface (MII). It is a standard interface for Ethernet device defined by IEEE-802.3 specification [10]. Based on this readout scheme, data can be easily sent to DAQ computer Ethernet connection.

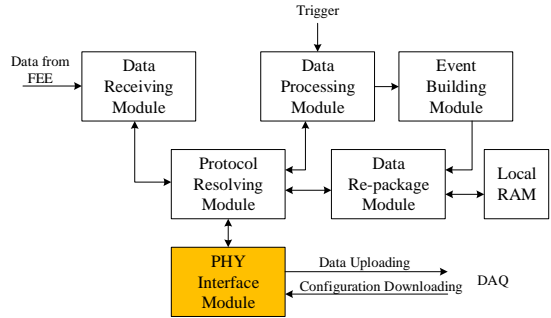

Fig. 4. (color online) FPGA logic Structure.

The FPGA logic architecture is shown in Fig. 4. Data receiving module controls the interface between readout module and FEE. The raw data from FEE can be obtained by the protocol resolving module. This protocol is customized between FEE and readout system for the sake of data transmitting correctly, efficiently and reliably. According to the absence or existence of trigger signal, data processing module decides whether the received data is valid or not. The data zero compression algorithm can also be implemented in this module. After being processed, valid data is transmitted to event building module, which will package the raw valid data according physics experiment requirements. Finally, for the purpose of data transmission over Ethernet PHY, the data should be re-packaged and buffered in local RAM waiting for being read out. The data packet format is shown in table 1.

Module No. refers to the number of each readout module. It can be configured during system initializing. With the help of this number, DAQ can construct the relationship between readout module and FEE channel.

Counting No. refers to the packet number in data flow. Obviously, the whole data flow is divided into many fragments. One counting number refers to one fragment. Once one data fragment packet missed or does not pass the correctness check, DAQ can require readout module retrieve and re-transmit the corresponding packet according to this number.

Packet size refers to the actual size of raw data in one packet. For the sake of simplicity and efficiency of transmission, data packet size is designed into a fixed size. It is 1024 in this paper. For the case of one long size data flow (>1024), the last data fragment size may be little than 1024, e.g. 500. The packet size should be 500 and the rest data in this packet should be 0.

Raw data refers to the valid data waiting for being transmitted to DAQ.

The CRC item refers to the checking data according cyclic redundancy check algorithm. Once DAQ finds out CRC data mismatch occurs, it sends data control command to readout module for re-transmitting the wrong data fragment.

The PHY interface module is the data swap channel between readout module and the DAQ. It controls the Ethernet PHY to execute the transmitting or receiving function through MII or GMII interface defined by IEEE-802.3 protocol.

Table 1. Raw data packet format.

| Item | Size (Byte) |
|---|---|
| Module No. | 1 |
| Counting No. | 1 |
| Packet Size | 1 |
| Raw Data | … |
| CRC | 1 |

Fig. 5 shows the data transmission state transition chart.

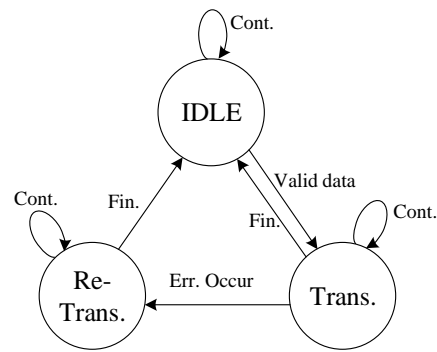

Fig. 5. Data transmission state transition chart.

The initial state for data transmission is idle state. In this state, the transmission module is waiting for valid data occurring. Once valid data is obtained and de-capsulated from FEE, the system enters into the transmission state. The duration for system residing in this state depends on the valid data size.

If there's no error occurs, system enters back into idle state to finish the transmission procedure. On the contrary, once any error (checking data mismatch) occurs, system enters into re-transmission state. The corresponding wrong data fragments will be re-transmitted to DAQ.

From the readout module side, the real-time performance and transmission efficiency can be easily guaranteed because the whole data path is implemented by full hardware without TCP/IP protocol and because the transmitted data over Ethernet is nearly all the raw data from FEE with only few redundant information. However, from the side of DAQ, the data packets from readout module cannot be recognized correctly because of the absence of standard TCP/IP protocol. To solve this problem, the standard data flow of an ordinary network interface card (NIC) should be modified.

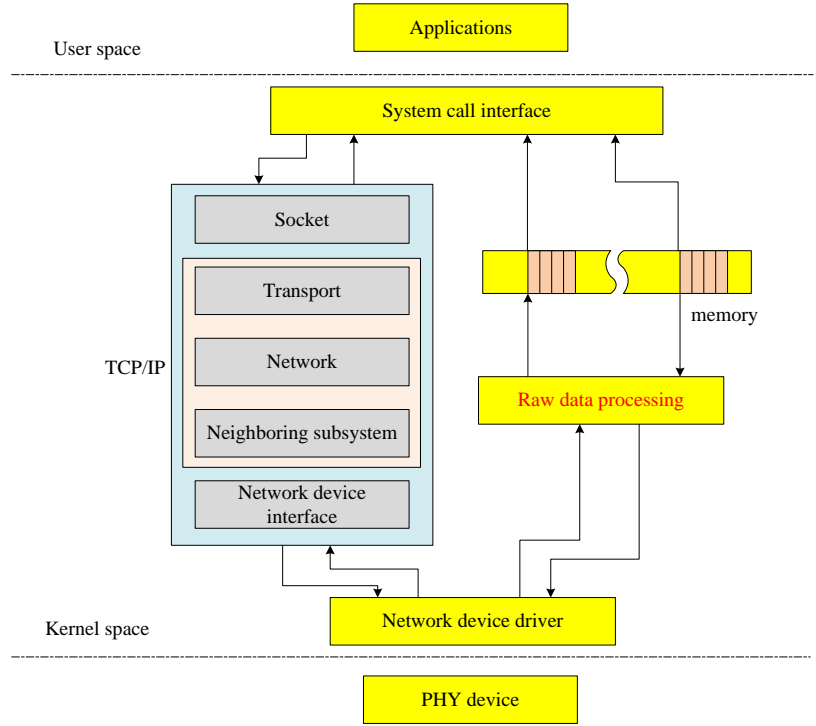

Fig. 6.  (color online) Data flow redirection.

In Linux operation system kernel space, the network device driver receives the data from PHY and sends to the TCP/IP protocol stack processing module. Generally, only after flowing through all the protocol layers one by one, data can be resolved correctly and can be sent to applications in user space eventually, as shown in the left side of Fig. 6.

To simplify the whole system design maximally, the easiest way to make the DAQ applications in user space recognize the received data flow correctly is to redirect the flow making the TCP/IP processing module being bypassed. As shown in the right side in Fig. 6, in the operating system kernel space, the data is redirected to a new path. The raw data processing module will de-capsulate and process the data according to customized protocol and then buffer the data into a memory which is mapped to user space during initialization state. So, once data is de-capsulated and stored into these buffers, user applications can access it directly. Now, from the view of device driver, the network card acts as a normal character device, not a networks device. The driver is simplified.

Although DAQ can communicate with the readout module with this modified device driver, however, when it communicates with other computers, it will be panic because of the problem of device driver compatibility. So the driver should also be able to recognize whether the data packet is coming from customized readout module or from normal computers.

Actually, Ethernet was designed before the IEEE created its 802.3 standard. The latter are not pure Ethernet, even though it is commonly called Ethernet standards. Fortunately, every Ethernet card is able to receive both the 802 standard frames and the old Ethernet frame. The main difference between these two kinds of network standard is that the header of Ethernet frame format as shown in Fig. 7. The number upon each item box refers to the size of this item in the frame [11].

To save space, the IEEE decided to use values greater than 1536 to represent the Ethernet protocol. The 802.3 protocol, however, use the field to store the length of the frame. This item is 2 bytes in size. The valid Ethernet types are predefined by IEEE and cannot be used arbitrarily. So, in this paper, we define the protocol type to be 0xFF00 that is different from the existed types. The data domain refers to the format defined in Tab. 1. The size is fixed to be 1024. Once network device driver receives packets with protocol type 0xFF00, it redirects them to the new customized raw data processing module in the kernel space. Obviously, by this way, the TCP/IP stack is bypassed.

| 6 | 6 | 2 | 0...1500 | 0..46 | 4 |
|---|---|---|---|---|---|
| Destination address | Source address | Protocol (>1536) | Data | Padding | Checksum |

(a) Ethernet frame format

| 6 | 6 | 2 | 0...1500 | 0..46 | 4 |
|---|---|---|---|---|---|
| Destination address | Source address | Length (<=1500) | Data | Padding | Checksum |

(b) 802.3 frame format

| 6 | 6 | 2 | 1024 | 0..46 | 4 |
|---|---|---|---|---|---|
| Destination address | Source address | Protocol (0xFF00) | Data | Padding | Checksum |

(c) Customized frame format

Fig. 7. Differences between Ethernet, 802.3 and customized frames.

## 3 Verification

To verify this proposed readout method, a prototype readout board is implemented as shown in Fig. 8. During verification, FPGA (a low cost Altera Cyclone chip) generates data and sends out through LVDS port for simulating the behavior of FEE. Then the data is looped through LVDS input port back to FPGA for simulating the behavior of readout module receiving data. The data is de-capsulated in FPGA and then sent to an Ethernet PHY chip (ANX5802, 100Mbps). Finally, PHY sends data to the DAQ computer.

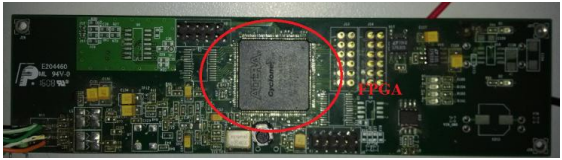
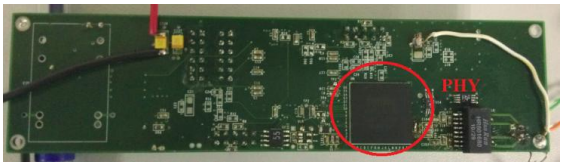

Fig. 8. (color online) Prototype readout board.

The test platform is shown in Fig. 9. Test result shows that this prototype of readout module can receive data from FEE and send to DAQ computer correctly and smoothly. The data throughput can reach up to 100Mbps.

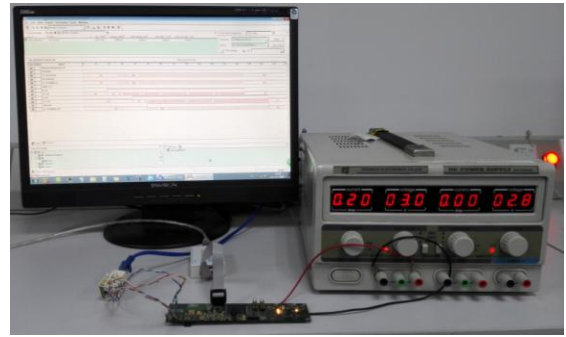

Fig. 9. (color online) Test platform.

## 4 Conclusions

In this paper, a novel and simple data readout method is proposed for the applications of physics experiment. The key point of this method is to try to obtain good simplicity and performance for Ethernet technique without TCP/IP protocol. The absence of TCP/IP protocol stack makes the readout module can be easily implemented, i.e. an FPGA combines with a PHY chip is enough for implementation. Simple and valid data re-transmission scheme can guarantee the reliability of data transmission. To make the non-standardized data packet be recognized correctly by DAQ software, a data flow redirection method is realized in operating system kernel space. To verify this method, a prototype readout board is implemented. Test result verified the feasibility and correctness of this method.

Besides physics experimental applications, this method can also be used in areas with the demands of data transmission over Ethernet.